\documentclass{article}[12pt]

\usepackage{fullpage}
\usepackage[hyphens]{url}
\usepackage{subfig}
\usepackage{multirow}
\usepackage{adjustbox}
\usepackage{graphicx,hyperref,fullpage,times,inconsolata,comment} 
    \hypersetup{colorlinks=true, breaklinks=true, linkcolor=blue, urlcolor=blue, citecolor=blue}

\usepackage[utf8]{inputenc}
\usepackage[english]{babel}
\usepackage{subcaption}
\usepackage[symbol*]{footmisc}
\usepackage[numbers]{natbib}
\usepackage{todonotes}
    \makeatletter
    \define@key{todonotes}{Mihai}[]{%
       \setkeys{todonotes}{inline,author={\textbf{Mihai}},color=green}%
    }
    \define@key{todonotes}{Somesh}[]{%
       \setkeys{todonotes}{inline,author={\textbf{Somesh}},color=orange}%
    }
\makeatother
\usepackage{booktabs}
\usepackage{tabularx}
\usepackage{tikz}
    \usetikzlibrary{positioning}
\usepackage{amsmath}
\usepackage{MnSymbol}
\usepackage{wasysym}

\usepackage{xcolor}

\title{Securing the Future of GenAI: Policy and Technology}
\author{
        Mihai Christodorescu \\ \textit{\small Google} \and
        Ryan Craven \\ \textit{\small Office of Naval Research} \and
        Soheil Feizi \\ \textit{\small University of Maryland, College Park} \and
        Neil Gong \\ \textit{\small Duke University} \and
        Mia Hoffmann \\ \textit{\small Georgetown University} \and
        Somesh Jha \\ \textit{\small University of Wisconsin, Madison and Google} \and
        Zhengyuan Jiang \\ \textit{\small Duke University} \and
        Mehrdad Saberi Kamarposhti \\ \textit{\small University of Maryland, College Park} \and
        John Mitchell \\ \textit{\small Stanford University} \and
        Jessica Newman \\ \textit{\small University of California, Berkeley} \and
        Emelia Probasco \\ \textit{\small Georgetown University} \and
        Yanjun Qi \\ \textit{\small University of Virginia} \and
        Khawaja Shams \\ \textit{\small Google} \and
        Matthew Turek \\ \textit{\small Defense Advanced Research Projects Agency}
}
\date{May 2024}

\begin{document}

\maketitle

\begin{abstract}
    The rise of Generative AI (GenAI) brings about transformative potential across sectors, but its dual-use nature also amplifies risks. Governments globally are grappling with the challenge of regulating GenAI, balancing innovation against safety. China, the United States (US), and the European Union (EU) are at the forefront with initiatives like the Management of Algorithmic Recommendations, the Executive Order, and the AI Act, respectively. However, the rapid evolution of GenAI capabilities often outpaces the development of comprehensive safety measures, creating a gap between regulatory needs and technical advancements.

    A workshop co-organized by Google, University of Wisconsin, Madison (UW-Madison), and Stanford University aimed to bridge this gap between GenAI policy and technology. The diverse stakeholders of the GenAI space---from the public and governments to academia and industry---make any safety measures under consideration more complex, as both technical feasibility and regulatory guidance must be realized. This paper summarizes the discussions during the workshop which addressed questions, such as: How regulation can be designed without hindering technological progress? How technology can evolve to meet regulatory standards? The interplay between legislation and technology is a very vast topic, and we don't claim that this paper is a comprehensive treatment on this topic. This paper is meant to capture findings based on the workshop, and hopefully, can guide discussion on this topic.

\end{abstract}

\section{Introduction}
\label{sec:intro}


The Cambrian explosion of Generative-AI (GenAI) capabilities and applications highlights the promise of broad impact GenAI holds in a variety of domains. At the same time, the dual-use characteristics of this technology introduce new risks and enhance existing risks. Governments around the world are taking note of the rapid expansion in terms of capabilities, applications, and risks and have introduced regulatory frameworks for GenAI, with the United States' Executive Order and the European Union's AI Act as some of the most recent such developments. Regulatory bodies are faced with a balancing act, where too-strict of regulation can stifle the technical development and economic growth of GenAI, while loose regulation can fail to provide sufficient guardrails around the impact of GenAI on society. Further complicating the matter is the fast pace of research and development in GenAI, where new technical capabilities are launched seemingly every day but are not yet comprehensive in terms of safety guarantees.

As an example, the US Executive Order on the Safe, Secure, and Trustworthy Development and Use of Artificial Intelligence, issued in October 2023~\cite{the_white_house_executive_2023}, highlights a number of areas at high risk due to the application of GenAI, ranging from nuclear, biological, and chemical weapons, to critical infrastructure and energy security, and financial stability and information fraud and deception. To this end, the Executive Order calls for standard setting, risk evaluation, and education efforts. Unfortunately, the research and development efforts in GenAI fall short of providing the technical capabilities to support these regulatory approaches. This leaves a rather large gap between regulatory requirements and technical abilities.

\begin{figure}
   \centering
---\includegraphics[width=0.95\textwidth]{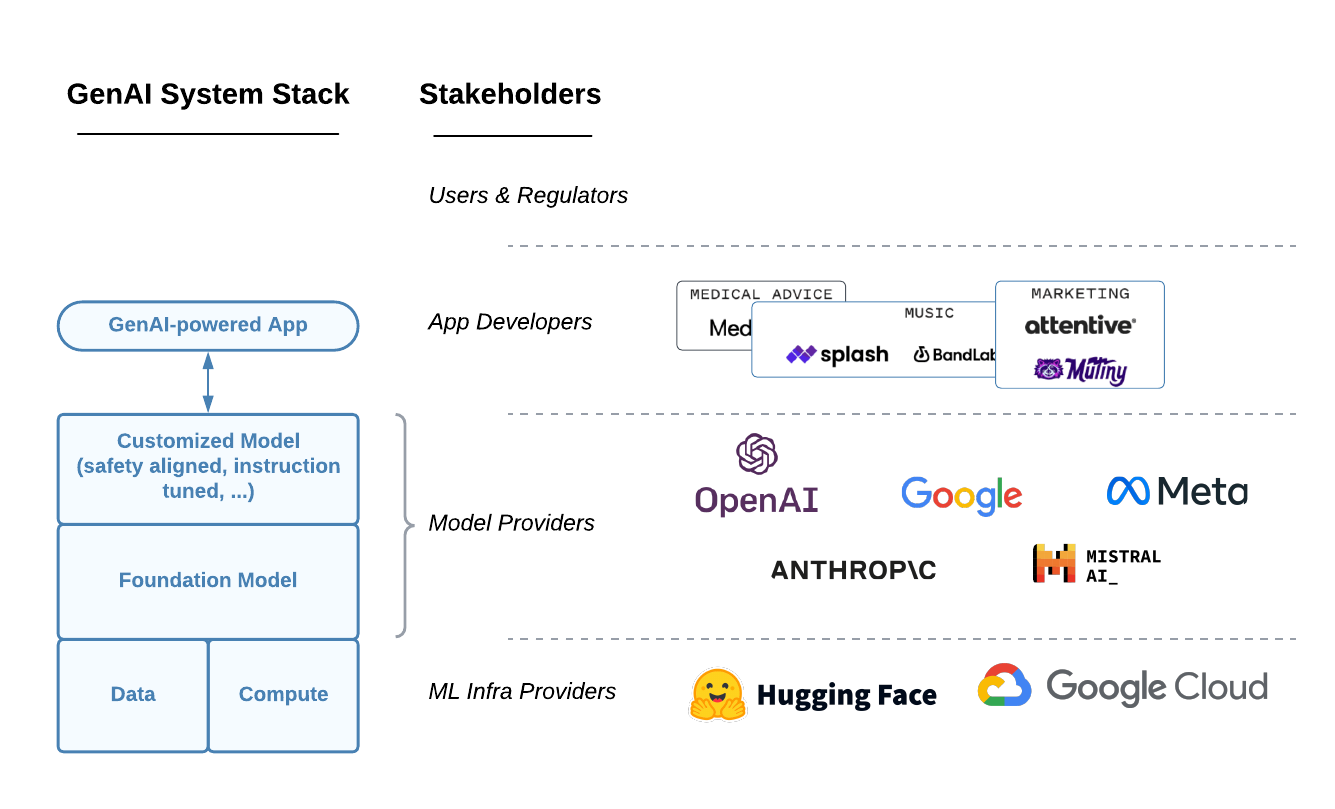}
   \caption{The software stack of GenAI-powered systems (shown here simplified to focus only on the components that can directly impact GenAI security) can have a variety of stakeholders, depending on distribution model. Data and compute providers have different leverage towards ensuring the security and safety of GenAI, compared to model providers and to app builders. Examples of GenAI apps were based on \url{https://www.sequoiacap.com/article/generative-ai-act-two/} .
   }
   \label{fig:llm-stakeholders}
\end{figure}

The GenAI policy and technology interplay is further complicated by the many stakeholders involved, from the general public as users of GenAI systems, to government agencies as regulators and enforcers of AI guardrails, and to academia and industry as technology creators. Within the technology space additional roles exist based on the GenAI development phase in which the stakeholders are involved, as the diagram in \autoref{fig:llm-stakeholders} illustrates. Thus any safety measures must not only take into account technical feasibility but also ownership and control of the GenAI model through its lifecycle, from pre-training, to fine-tuning, and to deployment.

We held a one-day workshop in October 2023 at Google~\cite{___securing_2023} to better understand the gap between GenAI regulation and GenAI technology, where a group of policy and technology experts convened to speak about their work. This workshop followed an earlier one in July 2023~\cite{___securing_2023-1} that considered the threats that GenAI creates or exacerbates and whose deliberations were published separately~\cite{barrett_identifying_2023}. The focus of the October workshop was at the intersection of regulatory policy and technology, in order to explore how regulation should be designed so as to guide technical evolution and how technology should be developed to meet regulatory requirements. This resulted in the following questions on the workshop shaping the agenda:
\begin{itemize}
    \item What are some important policy questions related to GenAI?
    \item What are the limits of GenAI safety alignment and is it achievable?
    \item What are limits of detecting whether content is GenAI generated?
\end{itemize}
This paper summarizes some of the findings and puts forward several goals for both GenAI policy makers and technology creators.

\vspace*{\baselineskip}
\noindent {\bf Detailed Roadmap.} \autoref{sec:policy} presents the landscape of approaches to GenAI regulation, both from individual governments and from multilateral governance bodies such as the Group of Seven (G7), as well as the lessons learned from military risk management. Subsequent sections discuss technological means to ensure the safety of GenAI models. First, we consider model alignment and its limitations in \autoref{sec:alignment}, then we look at model inspection (\autoref{sec:inspection}) and at provenance via GenAI output detection and watermarking (\autoref{sec:provenance}). In \autoref{sec:gaps} we summarize the gaps between the requirements of regulation and policymaking and the capabilities of current technologies for securing GenAI. Future directions and recommendations for both regulators and technologists are presented in \autoref{sec:future}, before the concluding remarks of \autoref{sec:conclusions}. We present this paper as the starting point for a discussion on how safety regulation and technical development of GenAI should progress and emphasize that it is not meant to be comprehensive. The focus is on summarizing the findings from the workshop and describing some interesting problems and challenges for the policy and technology communities.

\vspace*{\baselineskip}
\noindent {\bf Note.}
Given the nature of the topic, we welcome and value comments and feedback on our paper from the broader community.  We will address the feedback in future versions of the paper. Please send your comments and feedback to Mihai Christodorescu (\href{mailto:christodorescu@google.com}{christodorescu@google.com}), Somesh Jha (\href{mailto:jha@cs.wisc.edu}{jha@cs.wisc.edu}), or Khawaja Shams (\href{mailto:kshams@google.com}{kshams@google.com}).


\section{Regulatory-Policy Considerations for GenAI}
\label{sec:policy}


GenAI's seemingly broad impact across the whole range of human activities has attracted the attention of regulatory bodies in many countries in order to mitigate present and future risks of developing and deploying GenAI. A primary question is what types of GenAI uses and risks are of interest to regulators, and, relatedly, whether different regulators focus on different aspects of GenAI. In this section we summarize the workshop discussion on policies emerging in the European Union, the People's Republic of China, and the United States, as well as the governance efforts in multilateral settings (e.g., G7). Understanding the scope of such policies provides insights into the types of GenAI technologies that can be applied (or need to be developed) to achieve the required policies.

When GenAI falls short of perfect safety, it may be possible to design safeguards into the processes and procedures around using GenAI. The workshop participants found it useful to delve into how the (US) military addresses risk management in their own domain, where lethal technologies are omnipresent and without many safety options. In the last subsection we present a discussion of lessons from military risk management.

\subsection{The Policy Landscape}


Some governments around the world had made AI regulation a priority long before the release of OpenAI's ChatGPT in November 2022. For others, the sudden emergence of what seemed like a revolutionary technology reinvigorated interest in regulatory oversight. Policy responses emerged rapidly, and varied widely in terms of scope and subject. These differences are revealing. The approaches taken reflect how differently governments assess and prioritize the risks associated with GenAI.

Understanding the international policy landscape is important in its own right. In addition, insight into the underlying drivers of policy making by important geopolitical actors strengthens our ability to predict future action and identify pathways for international coordination and collaboration. A brief summary of the policy landscape across the European Union, People's Republic of China, and the United States is shown in \autoref{tab:landscape}.

\begin{table}
   \newcommand{\LE}{{\Large \textcolor{blue!30!white}{$\blacksquare$}}}
   \newcommand{\ME}{{\Large \textcolor{blue!30!white}{$\blacksquare$}\textcolor{blue!60!white}{$\blacksquare$}}}
   \newcommand{\HE}{{\Large \textcolor{blue!30!white}{$\blacksquare$}\textcolor{blue!60!white}{$\blacksquare$}\textcolor{blue!90!white}{$\blacksquare$}}}

   \caption{Various jurisdictions emphasize GenAI risks differently in their regulations. Legend: \LE\ = low emphasis, \ME\ = medium emphasis, \HE\ = high emphasis.}
   \label{tab:landscape}

   \centering
   \newcolumntype{Y}{>{\raggedright\arraybackslash}X}
   \begin{tabularx}{\textwidth}{lYYYYY}
      \toprule
      Jurisdiction   & Misuse & National Security and Competition & Data Integrity & Illegal Content & Algorithmic Harms \\
      \midrule
      European Union & \LE    & \LE                               & \HE            & \ME             & \HE               \\
      China          & \LE    & \HE                               & \ME            & \HE             & \LE               \\
      United States  & \HE    & \HE                               & \LE            & \ME             & \LE               \\
      \bottomrule
   \end{tabularx}
\end{table}

\paragraph{Scope.}

The analysis focuses on GenAI policy from the European Union (EU), the People’s Republic of China (PRC) and the United States (US), and is restricted to laws, regulations and proposals that directly pertain to GenAI. Broader policies that also apply to AI, like privacy regulation, are beyond the scope of this analysis, as are regulations from other countries and policies of commercial, for-profit organizations.

\paragraph{The PRC’s Interim Measures for the Management of GenAI.}

The Chinese government is an often overlooked first mover on AI regulation. It adopted rules for online recommender systems~\cite{china_law_translate_provisions_2021}
already in 2021. Regulation for AI-generated deep fakes~\cite{china_law_translate_provisions_2022}
followed in 2022, and another addressing generative language models for all modalities and concerning discrimination, bias, and intellectual property~\cite{china_law_translate_interim_2023}
was enacted in July 2023. 

The PRC’s GenAI regulation focuses on controlling content creation at the source. It aims to prevent the production and spread of content that fails to uphold core party values and violates the government’s strict censorship rules (the full definition of illegal and undesirable information can be found in the 2019 Provisions on the Governance of the Online Information Content Ecosystem order~\cite{zhang_provisions_2020}). To minimize the risk of incidents, the regulation sets out strict requirements on the provenance and composition of the training data. Providers of GenAI services must demonstrate their models’ compliance through a security self-assessment based on a detailed set of 
standards~\cite{chinese_national_information_security_standardization_technical_committee_sactc_260_basic_2024}, and are liable for deployed systems’ outputs. Whenever illegal content is generated, they must notify relevant authorities, and suspend and retrain or otherwise modify their system. Together with the regulation for recommender systems, which shapes how online content is promoted and consumed, the Interim Measures affirm the Chinese Communist Party’s commitment to controlling technology that can influence discourse and public opinion online.

At the same time, the regulation reveals the Chinese government’s careful balancing act between enforcing censorship and achieving AI leadership in a tense geopolitical environment. The draft legislative text underwent substantive amendments that resulted in a significantly relaxed final set of rules. It no longer contains some of the draft’s strictest requirements, such as guaranteed truthfulness of the training data and outputs and has several added articles promoting innovation and industry development. Most significant, however, is the narrowing of the regulation’s scope. Instead of covering all uses of GenAI, including research, as envisioned in the draft, the final rules only apply to public-facing GenAI services accessible in mainland China. This means that systems used in non-public contexts, for example in healthcare, public administration or industrial process automation are not subject to the regulation’s requirements. 

\paragraph{The European Union’s AI Act.}

On 8th December 2023 the EU reached an AI regulatory milestone. The political agreement between the European Parliament, the European Commission and the Council of the EU concludes years of preparation, development and negotiation on the most comprehensive regulatory framework on AI to date. The details of the final governance regime were published in May 2024 as the AI Act\footnote{\url{https://eur-lex.europa.eu/legal-content/EN/TXT/?uri=CONSIL\%3APE_24_2024_INIT&qid=1715851545785} }, with priorities going towards protecting people’s health, safety and fundamental rights from algorithmic harm through a tiered, risk-based regulatory framework.

Risk levels are assigned based on the use case in which AI systems are deployed. Those considered to pose unacceptable risks, such as manipulation and social scoring, are banned. Use cases believed to pose high risks are subject to the strictest safety and performance requirements in the Act, which include risk management, model evaluation, documentation, record-keeping and incident reporting, among other things. Examples of sensitive use cases are the areas of education, employment, public services and law enforcement. 

Rules for AI systems that do not serve a clearly defined purpose, such as GenAI systems powered by so-called foundation models, also follow a tiered system. All providers of general-purpose AI must publish information about their models’ training data and energy consumption, and label AI-generated content. Moreover, if their models are deployed in high-risk use cases, developers must provide the deployer with the documentation needed to comply with the corresponding regulatory requirements. In addition, models that pose systemic risks will need to comply with additional requirements, including model evaluation, risk management and cybersecurity protections. A general-purpose-model’s risk status is determined by the computing power used for training, the number of parameters and the number of business users. 

\paragraph{The United States Executive Order on AI.}

The release of ChatGPT accelerated and shifted AI policy discussions in Washington, D.C. Shortly before, the White House had published the 
Blueprint for an AI Bill of Rights~\cite{the_white_house_office_of_science_and_technology_blueprint_2022}, a set of voluntary guidelines for fairness, accountability and transparency in rights-impacting algorithmic decisions. The rise to popularity of GenAI refocused policymakers’ attention away from potential harms from AI’s use and towards emergent risk from AI models’ capabilities. 

Within one year, by October 2023, federal legislators held more than a dozen Congressional hearings on (generative) AI and introduced more than 50 
AI-related bills~\cite{brennan_center_for_justice_artificial_2024}. On 30 October 2023, President Biden signed the 
Executive Order on Safe, Secure and Trustworthy AI (EO)~\cite{the_white_house_executive_2023}, touching on a wide range of issues pertaining to AI safety, innovation, talent, civil and consumer rights, government use of AI, and more. The majority of legislative proposals, as well as the EO, focus on risks like deceptive content, emergent capabilities, and technology transfer.

The EO addresses the threat of AI-generated disinformation and weakened trust in government communications by investing in the development and use of content provenance and watermarking techniques. However, a second emphasis is placed on national security risks from AI. Advanced AI models might present with capabilities that pose new cyber, nuclear, biological or chemical risks. Adversaries innovating in and developing powerful AI models pose additional security threats. The EO therefore establishes a monitoring regime for advanced model development by domestic and foreign actors and directs significant resources to the development of model evaluation techniques to assess emergent capabilities and risks.

\paragraph{Divergent Priorities.}

Government approaches to GenAI governance are not developed in a vacuum. Instead, they are reflections of the societal, economic and political systems from which they emerge. The EU, as a technology importer, wields regulation as a lever to shape technology that is not designed in Europe. Having recognized the rights- and safety-impacting potential of AI early on, EU regulation addresses the impacts of simpler algorithmic systems equally alongside more advanced and capable models.

Conversely, the Chinese government’s approach reveals that the Chinese Communist Party not only views the ability of GenAI systems to produce potentially unaligned content at scale as their greatest threat, but also that innovation in the field is so critical to AI leadership that it is willing to compromise on its internal security policy to enable it. 
Finally, the US is a technological innovator amid a tense geopolitical environment in which technology plays an increasing role. This is reflected in an approach to AI governance that prioritizes control over technological development abroad, and supportive awareness of technological capabilities at home. While conventional algorithmic risks are recognized, addressing them would require the current Congressional gridlock be resolved.

GenAI developers wishing to serve all three markets must therefore find ways to meet the differing requirements for safety, transparency, and risk management that reflect these governments’ divergent priorities. Whether this can be achieved in light of the technical limitations that we describe in Sections~3--5 remains an open question.


\subsection{Multilateral Governance}
\label{sec:policy-multilateral}


In addition to the emerging regulatory environment for AI, there have been a growing number of multilateral governance efforts that specifically address the security risks of AI and generative AI. The G7 Hiroshima Process International Code of Conduct for Organizations Developing Advanced AI Systems (“G7 Code of Conduct”)~\cite{g7_hiroshima_2023} and The Bletchley Declaration~\cite{countries_attending_the_ai_safety_summit_bletchley_2023} are particularly notable for their emphasis on generative AI and general purpose AI compared to the broader landscape of multilateral AI governance. This more recent shift of attention has brought increasing specificity and urgency to the calls from multilateral governance bodies, and may more directly impact the actions of organizations around the world.

\vspace*{\baselineskip}\noindent\textbf{The G7 Code of Conduct} was published in October 2023. It specifically calls on the responsibility of developers of advanced AI systems to commit to follow the Code of Conduct. 

The Code of Conduct includes eleven actions for organizations to follow in a manner that is commensurate with the risks. The actions include identifying, evaluating, and mitigating risks across the AI lifecycle (including cyber capabilities, CBRN, “self-replication”, and risks to society, democracy, and human rights); identifying and mitigating vulnerabilities, incidents, and misuse after deployment (including implementing accessible reporting mechanisms and bounty systems to incentivize responsible disclosure of weaknesses); publicly reporting advanced AI systems’ capabilities, limitations, and domains of appropriate and inappropriate use (including the results of red-teaming); implementing robust security controls, including physical security and cybersecurity across the AI lifecycle; deploying reliable content authentication and provenance mechanisms to enable users to identify AI-generated content; and others. 

The relative specificity of the Code of Conduct in terms of who is responsible and what actions should be taken stands out in comparison to many AI principles and appears to be supporting expedient consideration. For example, in a November 2023 article from AI company Anthropic, the company stated, ``Anthropic supports the G7 Code of Conduct, which will inform our development and deployment practices, alongside the White House Commitments.''~\cite{anthropic_thoughts_2023}

\vspace*{\baselineskip}\noindent\textbf{The Bletchley Declaration} was published November 2023 following the inaugural UK AI Safety Summit. Twenty-eight countries from around the world, including the United States, China, Brazil, India, and Nigeria, agreed to the Declaration, as did the European Union. The Declaration calls for safe, human-centric, trustworthy, and responsible AI design, development, deployment, and use. Issues that are deemed critically important include the protection of human rights, transparency and explainability, fairness, accountability, regulation, safety, appropriate human oversight, ethics, bias mitigation, privacy and data protection, and manipulated or deceptive generated content. Except for the final issue regarding content generation, all of these issues have also been emphasized in earlier multilateral AI agreements such as the OECD Recommendation on Artificial Intelligence and AI Principles, later endorsed by the G20. 

The Bletchley Declaration stands out for additionally highlighting risks posed by “frontier” AI and “highly capable general-purpose AI models, including foundation models, that could perform a wide variety of tasks”. The risks described from these AI systems include both their misuse (for example to develop cyberattacks or biological weapons) and unintended risks stemming from a lack of understanding and control of an AI system. The Declaration affirms that developers of frontier AI systems should accept primary responsibility for ensuring their safety. 

The Declaration further calls for building a shared scientific and evidence-based understanding of these risks and building respective risk-based policies across countries, with a focus on transparency, evaluation, and testing. It also calls for greater international cooperation and dialogue, including an internationally inclusive network of scientific research on “frontier AI safety,” as well as support for developing countries to strengthen AI capacity building.
 
\begin{center} $\diamond$ \end{center}

\vspace*{\baselineskip}\noindent
The G7 Code of Conduct and the The Bletchley Declaration are a ``call to arms'' for both organizations and governments around the world and are a direct response to recent advances in general purpose and generative AI. However, they build on top of and interconnect with the broader landscape of multilateral AI governance, including the OECD Recommendation on Artificial Intelligence and AI Principles, the Global Partnership on Artificial Intelligence (GPAI), the UNESCO Recommendation on the Ethics of Artificial Intelligence, the UN Global Digital Compact, and China's Global AI Governance Initiative, among others. For example, the G7 Code of Conduct explicitly states that it builds on the OECD AI principles, which were first announced in 2019 and have been adopted by dozens of countries around the world, including by the G20 and numerous non-OECD countries. 

The US has historically favored working with the G7, OECD, and GPAI, as it has emphasized the importance of working with like-minded democracies of the world. In October 2023 China launched the Global AI Governance Initiative in part to offer an alternative path to the US, stressing its openness and inclusivity in contrast to the restrictions and export controls the US has leveraged against China. The inclusion and acceptance by China to participate in the UK AI Safety Summit and resulting Bletchley Declaration just one month later was thus highly uncertain. That there was agreement found in the Declaration marked a notable shift in the trajectory between China and the US and suggests that more inclusive multilateral forums that focus relatively narrowly on safety and security considerations may be possible. 

The UN also plays a critical role in international AI governance, in particular for the greater inclusivity of the Global South in AI governance deliberations. Work on AI governance at the UN has taken multiple forms, including the UN Global Digital Compact, which takes a broader view and outlines "shared principles for an open, free and secure digital future for all,” as well as the UNESCO Recommendation on the Ethics of Artificial Intelligence, which was adopted by all 193 Member States in November 2021 and is referred to as the “first-ever global standard on AI ethics.” The US rejoined UNESCO in July 2023, becoming its 194th member state. More recently, the UN established a High-Level Advisory Body on Artificial Intelligence, which is providing further guidance on international governance of AI and published an interim report in December 2023. 


\subsection{Learning from Military Risk Management}
\label{sec:policy-military}


The ideal path to fulfilling the requirements of regulatory frameworks is to create GenAI models and build GenAI systems that are completely safe and secure. As we argue in later sections, there are unfortunately technical impossibilities that prevent us from creating such completely safe and secure GenAI offerings. Furthermore malicious actors may want to hide unsafe behaviors inside GenAI models, for example through backdoors. Thus it is important to consider the bigger picture of how GenAI will be used and how risks from these use cases must be mitigated. The experience of the (U.S.) military in handling, applying, and deploying potentially dangerous technologies may be instructive and may lead to capable approaches to securing and safeguarding GenAI use cases.



A security guard is bound to get bored and miss a few things on a TV monitor over the course of an 8-hour shift, but when they’re paying attention they can assess a security threat quickly. By contrast, a computer vision algorithm will never be bored or inattentive, but it is unreliable at discerning a security threat from mere movement on the screen. Both humans and AI systems have strengths and weaknesses, and in an ideal world, new technologies will emerge that harness the strengths and minimize the weaknesses of both. Current conversations have been focused on the technical development of AI systems—which is important and necessary work—but focus is also needed on ways to mitigate AI risks by improving the knowledge and performance of human operators. Thankfully, the challenge of developing humans and organizations to employ advanced and safety-critical technologies is not new, and the tech-policy community can look to the U.S. military for some useful lessons learned.

The military is not the only organization that must manage safety-impacting machinery, but they are perhaps the most experienced and disciplined when it comes to creating and enforcing behavioral standards around powerful tech. Centuries of experience have refined the myriad ways in which professional militaries control lethal technologies. These range from the longstanding tradition of badges, ribbons, and medals that officers wear to identify their experiences and training, to organizational approaches like separately designating infantry from artillery units (an innovation in 1776), and later establishing armor units for tanks in World War I. In brief, \textit{the military is a leader in the art and science of risk management for lethal technologies}.

There is widespread awareness of how the military manages technical risks of weapons, such as the requirements process or operational test and evaluation, but it is harder to find a succinct description of all the efforts taken to mitigate risks through human intervention. These efforts are well known, but seldom thought of as a risk mitigation method for powerful technologies. These human, vice technological, interventions include qualifications regimes, the delineation of roles and responsibilities, a continuous cycle of exercise and assessment, and the promulgation of standardized doctrine, tactics, techniques, and procedures. None of these sorts of efforts are unique to the military, hospital systems also require qualifications and standard procedures, but the military has been a trailblazer that others (including healthcare) have imitated. Companies and governments thinking about AI governance today could imitate this approach as well.

At the heart of human-factor risk mitigation in the military is the qualification. For service members, qualifying is a process of demonstrating knowledge sufficient for a service member to be entrusted to operate a weapon. A qualified individual is recognized in personnel records, through public ceremonies, and sometimes even by special pins and badges worn on their uniforms. The difficulty of the qualification process and the type of recognition varies according to the risk of the technology. 

Qualification processes are widespread in the private sector as well, and so too could they be used to address the risks of AI systems. For example, just as teachers are qualified to instruct particular curricula in public schools, so too could they be qualified to use AI tools that would review student performance or provide instructional assistance. Just as doctors are qualified to perform certain specialized procedures and undergo a process to be granted privileges to perform these procedures at their respective medical centers, so too should they be qualified to use a large language model for medical record review properly. These qualifications acknowledge the potential utility of AI tools to improve outcomes, while simultaneously working to address the known weaknesses of AI tools through human intelligence.  

In the first instance, qualifications simply prepare the operator to best use the system and understand its weaknesses. A thoughtful qualifications process that results in a verifiable designation for the individual can also communicate the seriousness of their responsibilities with an AI system to the operator themselves but also to members of the public who are affected by the AI system’s decisions. Maintaining a qualification can also be made contingent on the successful completion of regular assessments to ensure qualified users are kept up to date with new AI developments—which will most certainly happen as the field continues to rapidly evolve.   

The second benefit of qualifications as a governance mechanism is their role in accountability processes. While an unqualified individual may claim ignorance, a qualified individual will be recognized with a responsibility commensurate with their knowledge. Furthermore, organizations can manage who is authorized to use what system for which purpose based on their qualification (in the military, this is called “roles and responsibilities”). Tracking qualifications within an organization or across the nation using identification numbers can also enable better organizational governance as well as researchers looking to develop new and improved technical or human factor risk reduction approaches for AI systems.

While qualifications feed into the management of individuals, it is important to recall that the military will also qualify units (like a ship) and even entire groups (like a carrier battle group). Unit- and group-level qualifications reinforce standards across individuals and ensure that errors do not emerge at the seams that exist between distinct but intertwined people and systems. An example in the military might be the qualification of an aircraft carrier to launch and recover jets. By qualifying the aircraft carrier the military ensures that all the qualified individuals—the captain, the helmsman, the flight deckhands, the pilots, and many others—understand how to work as a team to safely achieve the mission.  
 

While there is precedence for leveraging qualification regimes to mitigate the risks of certain technologies, the concept has yet to fully enter the AI governance debate. This may be because the technology has been mostly employed by expert users to date. These users are inherently familiar with the capabilities and limitations of AI systems, many of which they have developed. However, instances of less knowledgeable users inadvertently causing harm have already emerged~\cite{wang_how_2022}. As AI continues to proliferate, the risks posed by unqualified users or in complete management regimes will also grow.

Addressing this gap in AI risk management will be no easy task. The technology itself is still evolving and global standards for the technology—much less the human interfaces—are not yet set. But it would be a mistake to wait until the technology is “ready” to start the development of qualification regimes. Policy-makers, organizational experts, legal teams, and technical experts will need time to convene and develop learning objectives and accompanying materials, standard operating procedures, and operational techniques. They will also have to establish the organizations responsible for maintaining and administering qualifications and designate mechanisms by which those organizations will be funded and also held accountable. While this work may be less glamorous than developing breakthroughs in AI, it will be no less important to the broad application of AI.

As an example of such an approach, consider the practice of \textit{ML red-teaming}, which consists of evaluating a GenAI (or more generally any ML) model for robustness against a broad range of attacks and for alignment with desirable properties (factuality, fairness, etc.). Simply stating that a model was red-teamed successfully is insufficient, since it often depends how comprehensive the red-teaming evaluation was done. This leads to the need to have standardized benchmarks for red-teaming and to qualify the experts performing the red-teaming exercise, both to establish their credentials (e.g., red-teaming expertise in video models) and to ensure that one can determine whether a model was red-teamed and thus is presumably safe for their use case (e.g., text summarization). A red-teaming qualification would cover roles and responsibilities, continuous assessment of skills and expertise, and standardization of model assurance levels and procedures.



\section{Risk Mitigation through Model Alignment}
\label{sec:alignment}


GenAI model alignment is the process of training and tuning a model such that it always performs as desired. There
are multiple definitions of alignment, but Wikipedia provides the following definition for \textit{AI alignment}~\cite{wikipedia_contributors_ai_2024}:
\begin{quote}
    AI alignment research aims to steer AI systems towards humans' intended goals, preferences, or ethical principles. An AI system is considered aligned if it advances its intended objectives. A misaligned AI system pursues some objectives, but not the intended ones.
\end{quote}
Meanwhile, OpenAI describes the goal of their alignment efforts as follows~
\footnote{See \url{https://openai.com/index/our-approach-to-alignment-research/}}
\begin{quote}
    Our alignment research aims to make artificial general intelligence (AGI) aligned with human values and follow human intent. We take an iterative, empirical approach: by attempting to align highly capable AI systems, we can learn what works and what doesn’t, thus refining our ability to make AI systems safer and more aligned. Using scientific experiments, we study how alignment techniques scale and where they will break.
\end{quote}
The common themes in both the definitions are "human values" and "human intent".
Perfectly aligned models are by definition safe and secure, since they have been made \textit{by construction} to achieve the intended objectives in ways that satisfy human expectations. Achieving aligned models is a challenging and still open problem as we highlight in this section. One of the dimensions of alignment, is ``do no harm'' (for example, most commercial models will refuse to respond 
to the following prompt $p = \mbox{``Show me steps to make a bomb''}$). However, several researchers have shown how to transform 
a prompt $p$ encoding a harmful intent to another prompt $p'$ so that $p$ and $p'$ are semantically equivalent, but GenAI model 
will respond with an answer to $p'$~\cite{zou_universal_2023,chao_jailbreaking_2023,mehrotra_tree_2023}. Moreover, simple ``band aids'', such as adding guardrails to restrict the output, don't
seem to work~\cite{PRP}. In light of these attacks, we ponder the following question: {\it What are challenges in achieving alignment?} We elaborate on this question in the next section. 

\subsection{Challenges in Achieving Alignment}

Large language models (LLMs) are quickly becoming an integral part of the Internet infrastructure and software applications. LLMs are being used to create more powerful online search, help software developers write code, and even power chatbots that help with customer service. LLMs are being integrated with corporate databases and documents to enable powerful Retrieval Augmented Generation (RAG)~\cite{lewis_retrieval-augmented_2021} scenarios when LLMs are adapted to specific domains and use-cases. However, these scenarios in effect expose a new attack surface to potentially confidential and proprietary enterprise data. 

As the rapid evolution of AI-enabled chatbots continues and their deployment becomes more prevalent online and in business applications, the need to align them with human values and make them robust against adversarial attacks comes to the forefront. The identification and mitigation of a variety of risk factors, such as vulnerabilities, is the goal of pre-deployment testing and evaluation of LLM's. Reinforcement learning from human feedback (RLHF) combined with  Red Teaming~\cite{casper_explore_2023,ganguli_red_2022} are the primary techniques today for alignment and vulnerability discovery and mitigation, aiming to make the chatbot more resilient against prompt injections~\cite{deepmind_building_2022}.  These techniques include testing for traditional cybersecurity vulnerabilities, bias, and discrimination, generation of harmful content, privacy violations, and emergent characteristics of LLM's, as well as evaluations of larger societal impacts~\cite{kinniment_evaluating_2023,shevlane_model_2023}.

At the same time, the race between model developers and their adversaries has begun, and both sides
are making great progress. There are no signs of abating in this race, which brings up the question about the long term equilibrium state: is it going to be a state of safety and stability or a condition similar to cybersecurity? 

Recent theoretical results show that the modern technique of using guardrails to enforce alignment and resist prompt injections is inherently not robust - there are theoretical limits on rigorous LLM censorship~\cite{glukhov_llm_2023}. Employing other means of mitigating such risks, e.g., setting up controlled model gateways and other cybersecurity mechanisms, are needed. In addition, adapting chatbots to downstream use-cases often involves the customization of the pre-trained LLM through further fine-tuning, which introduces new safety risks that  may degrade the safety alignment of the LLM~\cite{qi_fine-tuning_2023}.

Adversarial samples may be out-of-distribution (OOD) inputs. Thus, detecting OOD inputs is an important challenge in adversarial machine learning, and might help with attacks on alignment. Fang et al.~\cite{fang_is_2022} established  theoretical bounds on OOD detectability, i.e., an impossibility to detect when there is an overlap between the in-distribution and OOD data. 

As models grow in size, the amount of training data grows proportionally. Very few of the LLMs in use today publish a detailed list of the data sources used in training. Those that do~\cite{mitchell_bigscience_2022,touvron_llama_2023-1} show the scale of the footprint and the massive amounts of data consumed in training. The multi-modal generative AI systems exacerbate the demand further by requiring large amounts of data for each modality. 

Data repositories are not monolithic data containers but a list of labels and data links to other servers that actually contain the corresponding data samples. This creates new hard-to-mitigate risks~\cite{carlini_poisoning_2023}. In addition open source data poisoning tools~\cite{melissa_heikkila_this_2023} increase the risk of large scale attacks on image training data. 

Another scale-related problem is the ability to generate synthetic content at scale on the internet. Although watermarking may alleviate the situation, the existence of powerful open or ungoverned models creates realistic opportunities to generate massive amounts of unmarked synthetic content that can have a negative impact on the capabilities of subsequently trained LLMs~\cite{shumailov_curse_2023}, leading to model collapse. 
       
Based on this, one may conclude we are likely to land in a state similar to where cybersecurity is today.
Barrett et al.~\cite{barrett_uc_2023} have developed detailed risk profiles for cutting-edge generative AI systems that map well to the NIST AI RMF~\cite{national_institute_of_standards_and_technology_artificail_2023} and should be used for assessing and mitigating potentially catastrophic risks to society that may arise from this technology.


\section{Risk Mitigation through Model Inspection}
\label{sec:inspection}



Model inspection is key to ensuring the effectiveness, fairness, reliability, and transparency of generative AI systems. Model inspection includes a wide range of tasks, from using model interpretation methods to discovering the biases in language and image models~\cite{lin_word-level_2023,han_is_2023,wu_stable_2023} to novel adversarial attacks and safety evaluation frameworks~\cite{shayegani_survey_2023,zou_universal_2023}, offering insights into the multi-faceted nature of generative AI development and application. This body of work serves as a crucial resource for anyone interested in the ethical and technical dimensions of AI and machine learning.

\paragraph{Use Cases.}
By inspecting models, developers can ensure that a model produces reliable and trustworthy outputs, which is crucial for applications where accuracy, precision and/or safety are critical. One recent study from~\cite{weidinger_sociotechnical_2023} presents a framework for evaluating the safety of generative AI systems, highlighting the importance of integrating sociotechnical perspectives in AI safety evaluations. It outlines a three-layered approach: capability evaluation, human interaction evaluation, and systemic impact evaluation. This framework emphasizes the need for comprehensive safety assessments that consider technical aspects, human interactions, and broader systemic impacts. The paper also reviews the current state of safety evaluations for generative AI, identifying gaps and proposing steps to address them.

Model inspection methods also  allow for the identification of biases in a model's outputs. This is important for ensuring fairness and preventing discrimination in AI-generated content. A recent study~\cite{lin_word-level_2023} addresses the issue of under-representation in text-to-image stable-diffusion models. It introduces a method for identifying which words in input prompts contribute to biases in generated images. Their experiments show how specific words influence the replication of societal stereotypes. The paper also proposes a word-influence metric to guide practitioners in modifying prompts for more equitable representations, emphasizing the importance of addressing bias in AI models to prevent discrimination and stereotype perpetuation.
Another related study~\cite{wu_stable_2023} examines biases in text-to-image models, specifically Stable Diffusion. It introduces an evaluation protocol to analyze the impact of gender indicators on the generated images. The study reveals how gender indicators influence not only gender representation in images but also the depiction of objects and layouts. It also finds that neutral prompts tend to produce images more aligned with masculine prompts than feminine ones, providing insights into the nuanced gender biases in Stable Diffusion.

\paragraph{Inner Interpretability and Outer Explainability.}
Model inspection consists of a number of techniques meant to generate explanations or interpretations of a GenAI model's operation and outputs. Borrowing from Doshi-Velez and Kim~\cite{doshi-velez_towards_2017}, we use the following definition:
\begin{quote}
[ML system] interpretability [is] the ability to explain or to present in understandable terms to a human
\end{quote}
Depending on \textit{what} is being interpreted, we arrive at two classes of model-inspection techniques. If the goal is to relate the outputs of the model to the relevant parts of its inputs, then \textit{outer explainability} methods are applicable. If the goal is to relate the outputs of the model to the relevant inner structure (e.g., model weights, neurons, or subnetworks), then \textit{inner interpretability} methods are applicable. We note that both classes of techniques may rely on model internals (weights, gradients, layers) to achieve their goals.

\paragraph{Outer Explainability.}
Deep-learning literature includes a large cohort of different model interpretability methods in deep learning, including methods from saliency maps,  activation maximization,  layer-wise relevance propagation, partial dependence plots, LIME, SHAP, and Integrated Gradients, and more (see, for example, survey in \url{https://arxiv.org/abs/2011.07876}). These methods help to demystify the decision-making processes of deep learning models, making them more transparent and trustworthy. For instance,  the method PRIME~\cite{rezaei2023prime} proposes a new method for analyzing failure modes in image classification models using human-understandable concepts (tags) for images in the dataset and analyzing model behavior based on these tags. The method ensures that the tags describing a failure mode form a minimal set, avoiding redundant and noisy descriptions. Experiments demonstrate that this approach successfully identifies failure modes and generates high-quality text descriptions, emphasizing the importance of prioritizing interpretability in understanding model failures. In another study, authors of~\cite{wu_analyzing_2023} investigate how Chain-of-Thought (CoT) prompting affects LLMs. It examines whether CoT affects the importance given to specific input tokens by LLMs. Using gradient-based feature attribution methods, this study analyzes several open-source LLMs to understand changes in token importance due to CoT prompting. The findings indicate that while CoT doesn't increase the saliency scores of relevant tokens, it does enhance the robustness of these scores to variations in question phrasing and model outputs. This research provides insights into how CoT prompting influences LLM behavior, particularly in question-answering tasks.

Being able to explain model outputs is a key requirement especially for GenAI models whose complex architectures prevent the use of pre-Deep Neural Network explanatory methods and whose emergent behaviors are, by definition, not anticipated at training time~\cite{wei_emergent_2022}. \textit{Self explanations} are an interesting emergent behavior~\cite{wei_emergent_2022, chen_models_2023}, in which a model not only generates an output appropriate to the task given in the input, but can also explain its decisions in a manner understandable to humans. Unfortunately, self explanations turn out to be unreliable, failing to match the gradients of the model output~\cite{agarwal_faithfulness_2024}. Even the widely used and studied CoT technique can be manipulated into producing incorrect results together with highly confident explanations~\cite{turpin_language_2023}.

\paragraph{Inner Interpretability.}
Most studies on model interpretability focus on providing  post-hoc explanations that identify features or feature interactions that contribute most to a model's predictions.  Recent literature has shown increasing interest in treating interpretability as an inherent property of deep learning models or using interpretations as feedback to improve model performance and encourage explanation faithfulness.

We discuss here \textit{mechanistic interpretability}, a particular flavor of interpretability that focuses on identifying specific components of the neural network, such as individual neurons or groups of neurons or subnetworks/circuits, whose operation is responsible for detecting particular properties of the input or guaranteeing particular properties of the output. For example, analysis of the Inception-V1 vision model uncovered neurons that detect curves in images, or detect image patches at the boundary of high-frequency and low-frequency regions, or detect dog heads~\cite{olah_zoom_2020}. Performing such an analysis over the whole neural network of a model may be able to enumerate over all ``features'' of a model and then using that information to establish that a model is safe because all of its component features are safe and desirable for the task at hand, or, vice-versa, that a model is unsafe because one or more of its component features are known to be harmful or simply have unknown behavior. Anthropic's paper ``A Toy Model of Superposition''~\cite{elhage_toy_2022} posits that mechanistic interpretability applied at scale may be the path to establishing model safety through feature decomposition, an approach that came to be known as \textit{enumerative safety}.

The same paper~\cite{elhage_toy_2022} also points out that neurons are not always related to a single property of the input or the output, but some rather operate to capture multiple, often unrelated properties. These polysemantic neurons, possibly brought forth by the superposition hypothesis~\cite{goh_decoding_nodate}, complicate the use of mechanistic interpretability to establish enumerative safety. At a minimum, polysemantic neurons increase the cost of enumerating all potential components of a model, since now components are no longer disjoint as they may share neurons. A second dimension that increases the complexity of enumerative safety is the observation that (linear) groups of neurons form better unitary components in a model than individual neurons~\cite{bricken_towards_2023}. So enumerative-safety approaches must consider both individual neurons and groups of neurons. While promising, the science of mechanistic interpretability is still in its infancy and has been shown to work only for toy models, nowhere close in size to the large models available today.


Another important topic in model inspection is the evaluation of model interpretation. In deep learning, validating the faithfulness of model interpretability methods is crucial for ensuring that the decisions made by AI systems are understood correctly, trusted, and aligned with human values and societal norms. The literature includes two fast-growing research groups of evaluation strategies on model explanations: automatic (objective) evaluation and human (subjective) evaluation~\cite{carvalho2019machine}. Human evaluation methods identify whether a generated interpretation is useful for human users to understand model predictions. Literature includes many automated strategies to quantify the quality of model explanations. For instance, \cite{tanneru_quantifying_2023} focuses on measuring the uncertainty in explanations provided by LLMs. It introduces two new metrics, Verbalized Uncertainty, and Probing Uncertainty, to assess the reliability of explanations generated by LLMs. This study reveals that verbalized uncertainty is not a reliable estimate of explanation confidence while probing uncertainty correlates with the faithfulness of an explanation. The paper discusses the significance of understanding the uncertainty in LLM explanations to ensure trustworthiness and avoid plausible but inaccurate explanations. This research contributes to enhancing the transparency and reliability of foundational models in natural language processing. 

Broadly speaking, adversarial attack methods are also doing model inspection. An adversarial attack against a LLM like GPT-3 used carefully designed strategies to deliberately trick the model into making errors or producing unintended responses. These attacks exploit weaknesses or blind spots in a model's understanding of  its underlying data and  algorithms. Understanding and defending against  adversarial attacks is crucial for the responsible development and deployment of AI, especially in areas where accurate, safe and unbiased AI responses are critical. Our workshop includes one presentation from a notable recent LLM attack from \cite{zou_universal_2023} that conducts adversarial attacks on aligned LLMs. It focuses on generating objectionable content by appending a specially crafted suffix to various user queries. This approach combines greedy and gradient-based optimization techniques to optimize these adversarial suffixes, making them effective across different models and prompts. The study demonstrates that these attacks are highly transferable, even to black-box, publicly released production LLMs. The results significantly advance the state-of-the-art in adversarial attacks against LLMs, raising important questions about the robustness and safety of these AI systems. Recent literature on LLM adversarial attacks has been burgeoning, primarily due to the growing integration of LLM models into various real-world applications (surveyed in \cite{shayegani_survey_2023}). This line of research is crucial as it sheds light on the vulnerabilities of LLMs and helps in developing robust, secure generative AI systems.


\begin{figure}
   \centering
   \includegraphics[width=0.65\textwidth]{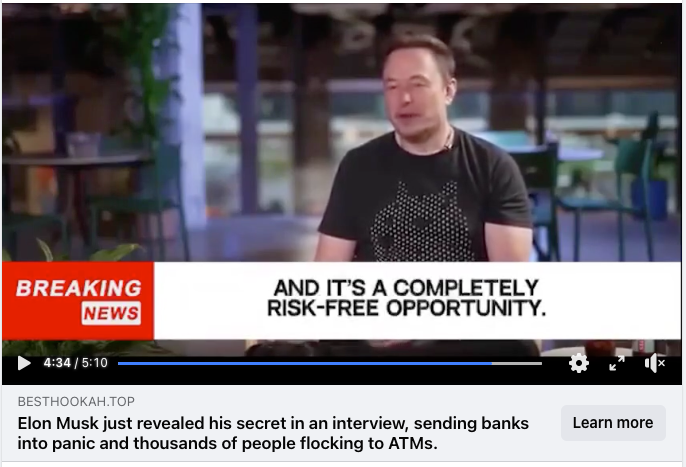}
   \caption{Deepfakes can be used to promote investment scams. This screenshot is from a deepfake video that circulated in November 2023 on social media, primarily targetting South African users, in which Bongiwe Zwane and Francis Herd from the South African Broadcasting Corporation (SABC, South Africa’s public TV and radio broadcaster) and Elon Musk appeared to promote an investment opportunity. The video is staged as a news clip, with a brief introduction from (deepfake) Herd, followed by (deepfake) Musk announcing ``powerful, world-first investment software'' while on a stage. SABC, Zwane, and Herd all denounced the deepfake through their web and social media presence (see \url{https://www.sabcnews.com/sabcnews/894115-2/}, \url{https://www.facebook.com/bongiwe.khumalo.946/videos/1151663159551664}, and \url{https://twitter.com/FrancisHerd/status/1721835389994799321}). Screenshot and details from \url{https://africacheck.org/fact-checks/meta-programme-fact-checks/beware-another-elon-musk-investment-scam-using-deepfake} .
            }
   \label{fig:taylor-trump-fake}
\end{figure}

\section{Risk Mitigation through Provenance and Watermarking}
\label{sec:provenance}

One of the central problems in the era of GenAI is {\it provenance tracking} or the ``GenAI Turing Test (GTT)'' (e.g., was  content $x$ generated by a known GenAI system (Claude, GPT, DALL-E, Gemini) or a natural image). Recall that {\it deepfakes} are synthetic media that have been digitally manipulated to replace one person's likeness convincingly with that of another or to place a person convincingly in a fake setting (an example of a deepfake is shown in \autoref{fig:taylor-trump-fake}). Currently, deepfakes are mostly generated using GenAI techniques, so provenance tracking and deepfake detection are very closely related problem, but not exactly the same (deepfakes can be generated without using GenAI techniques).

Given the importance of deepfake detection, attribution, and mitigating techniques, such as watermarking~\cite{jiang2024watermark}, several laws related to AI prominently mention these topics. For example, the executive order from the White House mentions the 
following\footnote{\url{https://www.whitehouse.gov/briefing-room/statements-releases/2023/10/30/fact-sheet-president-biden-issues-executive-order-on-safe-secure-and-trustworthy-artificial-intelligence/}}:
\begin{quote}
Protect Americans from AI-enabled fraud and deception by establishing standards and best practices for detecting AI-generated content and authenticating official content. The Department of Commerce will develop guidance for content authentication and watermarking to clearly label AI-generated content. Federal agencies will use these tools to make it easy for Americans to know that the communications they receive from their government are authentic—and set an example for the private sector and governments around the world.
\end{quote} 
Recently, Prime Minister Narendra Modi advocated the use of watermarks on AI-generated content to curb misinformation and deepfake-related harms in society\footnote{\url{https://www.newindianexpress.com/business/2024/Mar/30/use-watermarks-on-ai-content-pm}}.

Deepfake detection is discussed in \autoref{sec:medifor+semafor}. Watermarking is a promising technique for tracking provenance of GenAI content and is discussed in \autoref{sec:watermarking}.



\subsection{Detection of AI-Generated Output}
\label{sec:medifor+semafor}

\begin{figure}
   \centering
   \subfloat[position=top][Three images from 1864 show, clockwise from top left, prisoners from the Battle at Fisher's Hill (Va.), Gen. Ulysses S. Grant at headquarters in Cold Harbor (Va.), and Maj. Gen. Alexander McDowell McCook.]%
      {\adjustbox{valign=b}{\begin{tabular}{@{}c@{\ }c@{}}
          \includegraphics[width=0.22\textwidth]{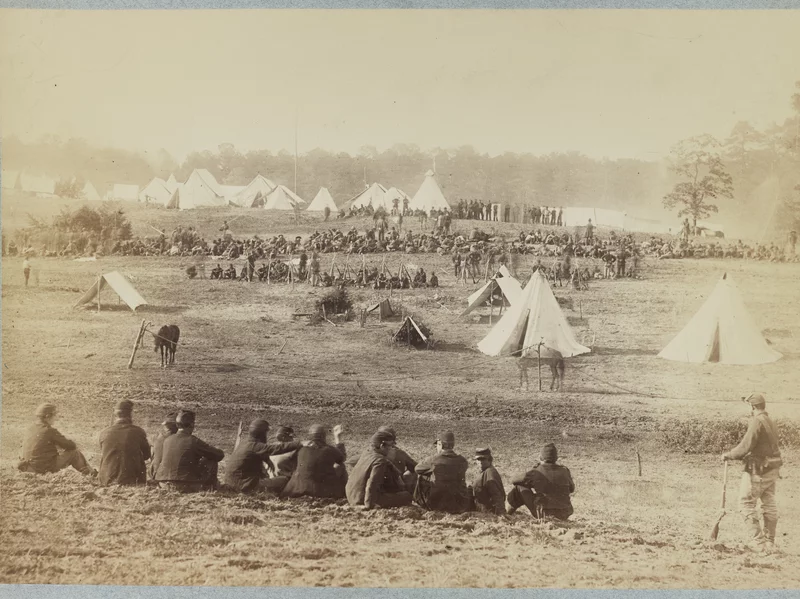}
          & \\
          \includegraphics[width=0.22\textwidth]{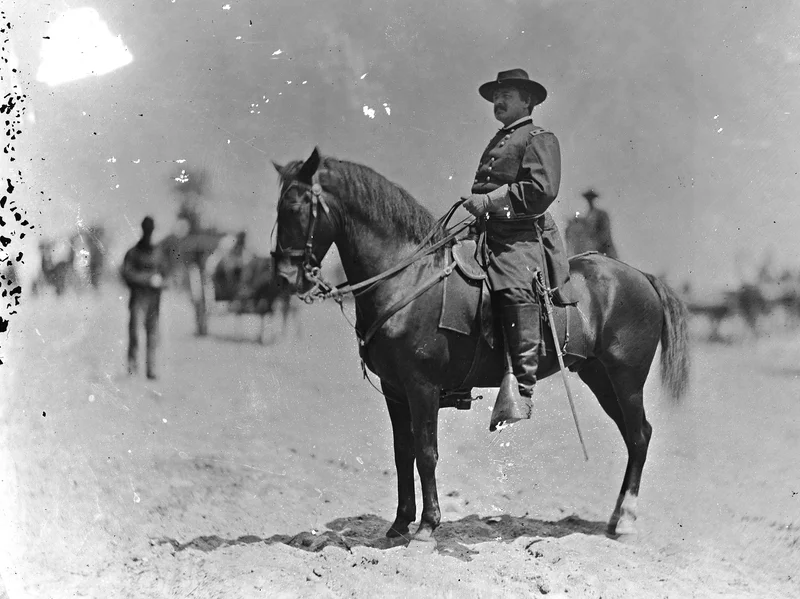}
          & \multirow[t]{2}{*}{\includegraphics[width=0.25325\textwidth]{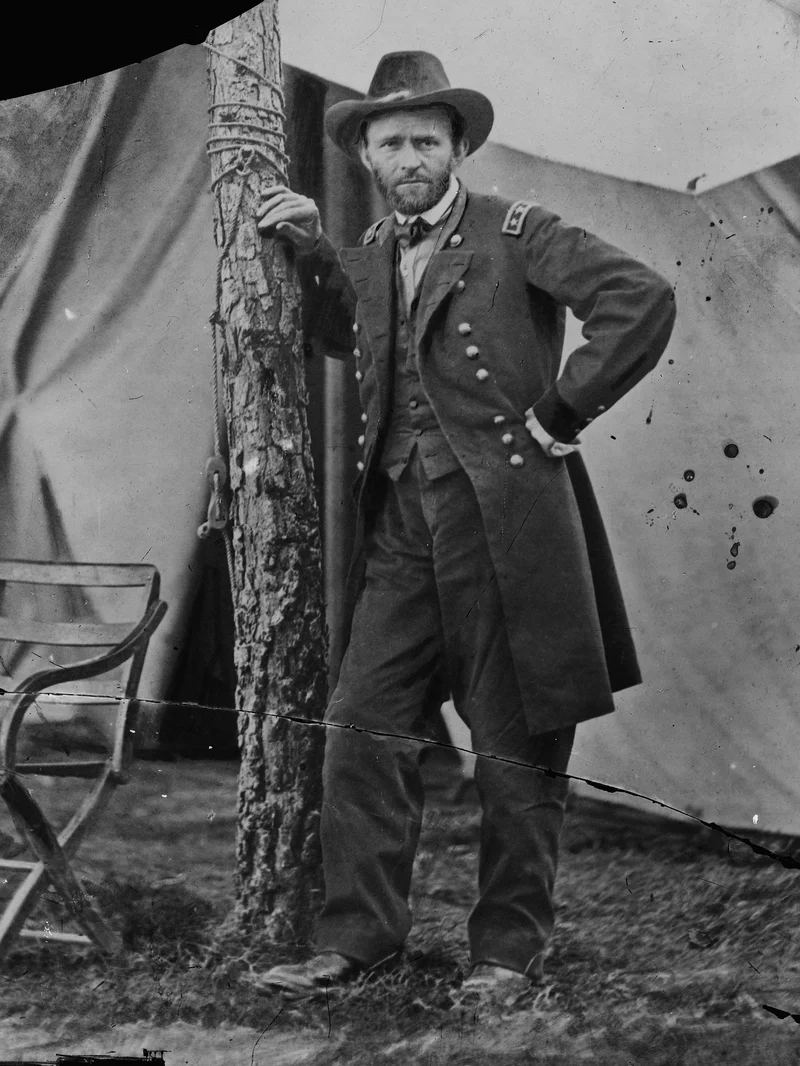}}
       \end{tabular}
       }%
      }
   \quad
   \subfloat[position=top][Fake image of General Grant at City Point, combined from the three images at left, perhaps by L.C. Handy circa 1902.]%
      {\adjustbox{valign=b}{\begin{tabular}{@{}c@{}} \includegraphics[width=0.4525\textwidth]{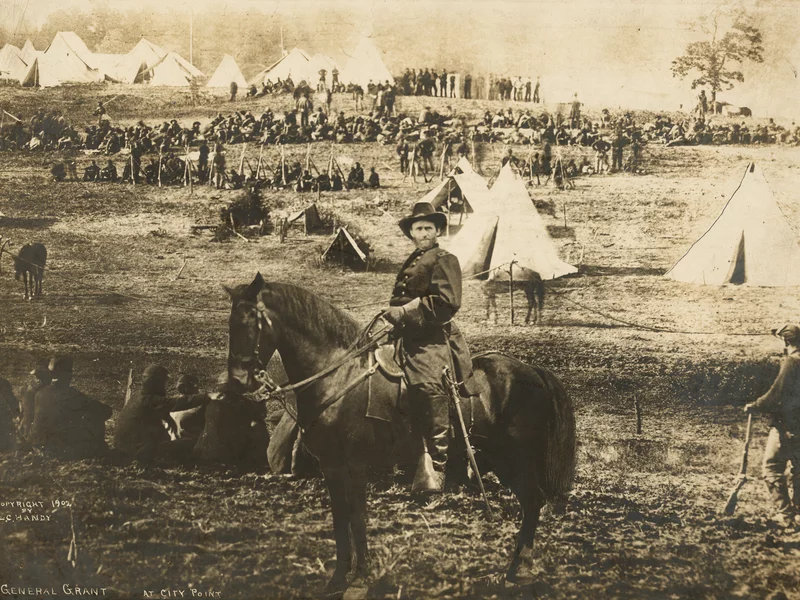} \end{tabular} } }
   \caption{Media manipulation has been happening long before GenAI. In the above example from 1902, three Civil War photos were combined to create a fake image of General Ulysses S. Grant. Images from Library of Congress, information from \url{https://www.npr.org/sections/npr-history-dept/2015/10/27/452089384/a-very-weird-photo-of-ulysses-s-grant} . }
   \label{fig:civil-war-photo-fake}
\end{figure}

Media falsification has existed since the beginning of media (e.g., photographs as shown in~\autoref{fig:civil-war-photo-fake}). Early commercial photographers used darkroom techniques to create falsified images to monetize large collections of negatives~\cite{fineman_faking_2012}. Young photographers created falsified photographs that were published and fooled parts of the public using cardboard cutouts of fairies~\cite{holmes_cottingley_2012,museum_of_hoaxes_cottingley_nodate}. Stalin used an army of retouchers to help falsify the history portrayed in photographs~\cite{fineman_faking_2012,blakemore_how_2022}. These examples illustrate the use of media manipulation for profit, entertainment and attention, and as a political weapon. In the context of GenAI, what’s new is the lowering of the level of skills and resources necessary to create a compelling falsification. Lowering the bar to compelling falsification enables more potential adversaries and potentially a much larger scale of media-falsification attacks.

Several potential attack vectors have been identified over the last several years, such as Jordan Peele’s public service announcement demonstrating a deepfake of President Obama~\cite{mack_this_2018}; Bricman’s coining of the term Ransomfake~\cite{bricman_deepfake_2019}, a mashup of ransomware and deepfake, where generated media is used to put someone in a comprising position unless a ransom is paid; and large-scale generated events~\cite{turek_semantic_2019}. As of 2023, many of these sorts of attacks have been seen in the wild, particularly with deepfakes of President Zelenskyy and President Putin during the Russia-Ukraine war~\cite{demery_demeryuk_wondered_2022,reuters_fact_check_doctored_2022} and the generation of media purporting to be falsified historical events~\cite{paleja_fake_2023}. 

Given the gravity of this problem, the US Department of Defense (DoD) has invested significant resources in tackling this problem. In the context of these challenges, the Defense Advanced Research Projects Agency (DARPA) made two significant investments. The first was the {\it Media Forensics (MediFor)} program, from 2016 to 2020. MediFor sought to produce quantitative measures of media integrity for images and video to enable integrity assessment at scale. The second is the {\it Semantic Forensics (SemaFor)} program that seeks to create rich semantic algorithms that automatically detect, attribute, and characterize multi-modal media. We discuss this problem through the lens of these DARPA programs as many prominent research groups that work on this problem were part of this program, and these programs are producing groundbreaking techniques in the context of deepfake detection.

There are four key problems in media authentication. The first is {\it detection} – determining whether a media asset (image, video, audio, or text) is real, manipulated, or AI-generated. The second is {\it attribution} – determining the source of a media asset and whether that source is consistent with the purported source. The third is {\it characterization} – identifying a rationale or intent behind the manipulation. The fourth is producing {\it evidence} that supports detection, attribution, or characterization. Attribution can be particularly important from a US government perspective, as various legal authorities to respond may be contingent on the actor behind the falsified media. Characterization is important as AI-generated media becomes commonplace and perhaps predominant and the challenge is to help surface media generated for malicious purposes.

MediFor and SemaFor have leveraged three categories of integrity, namely digital, physical, and semantic integrity. Digital integrity refers to digital artifacts left behind by authentic or inauthentic media processes, and may include compression artifacts, photoresponse non-uniformities (PRNU), and high frequency artifacts, to name a few~\cite{farid_photo_2016}. Physical integrity looks for indications that the laws of physics have been violated, for instance inconsistencies in scene geometry or lighting. Finally, semantic integrity looks for inconsistencies with respect to other sources of information, such as inconsistencies in weather based on time and location.

\begin{figure}
    \centering
    \includegraphics[width=\textwidth]{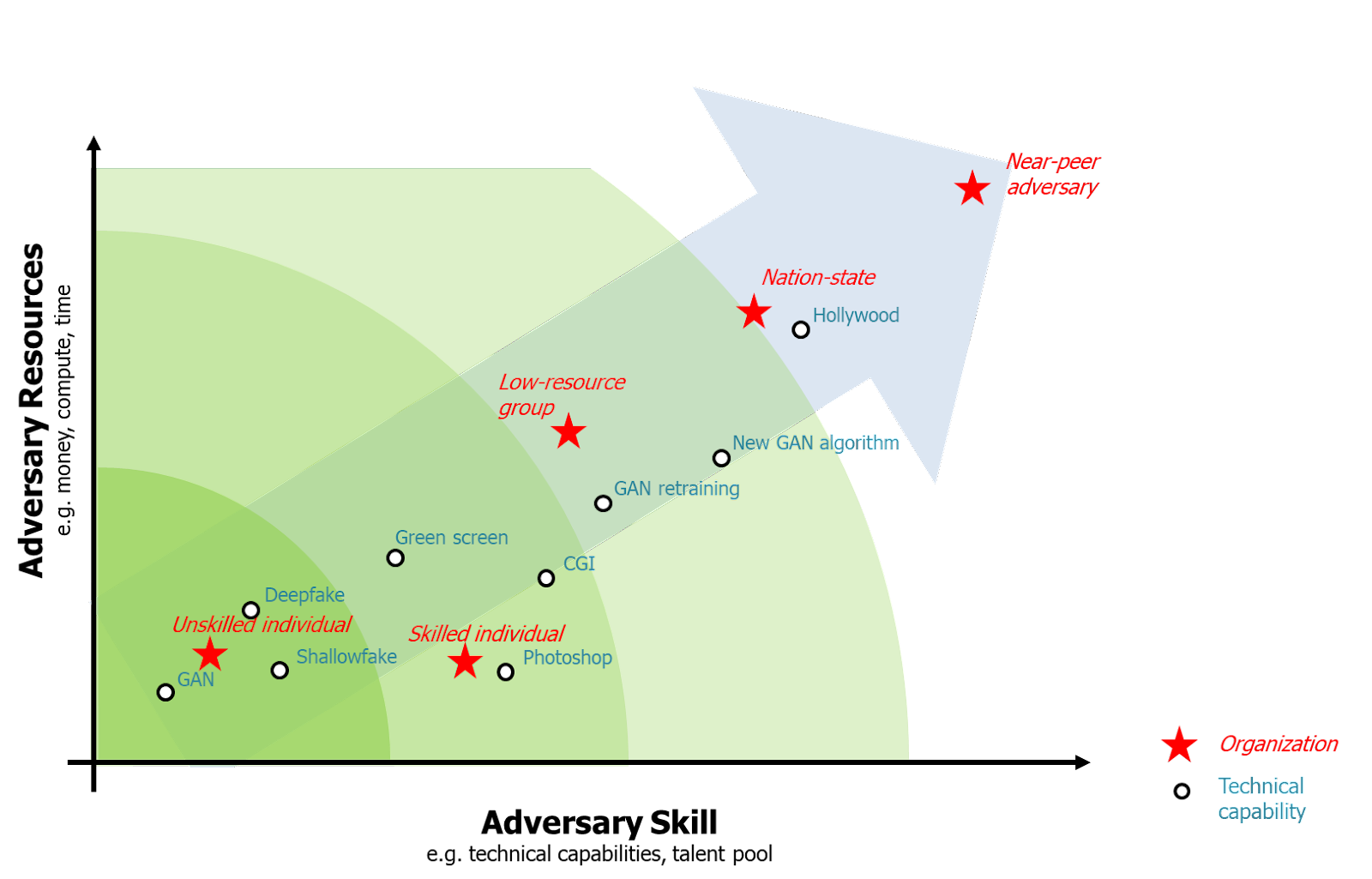}
    \caption{Notional adversarial landscape for media falsification. The x-axis shows various adversaries increasing in capability. The y-axis shows resources available to these classes of adversaries.}
    \label{fig:deepfake-landscape}
\end{figure}

It’s important to consider potential classes of adversaries in the context of media falsification. MediFor developed a notional plot of adversary skill and adversary resources necessary to leverage particular falsification techniques, as shown in~\autoref{fig:deepfake-landscape}. The threat landscape can be particularly useful to help understand what sort of falsification attack might come from various categories of adversaries. The analysis can also be useful to identify techniques where it might be important to deploy proactive defenses, like digital provenance and watermarking. SemaFor built on this approach, with a dedicated team to provide analysis of potential future threat landscapes. There is significant room for research that helps concretize and quantify potential future threats, e.g., economic models of potential threats.

Understanding of the threat landscape helped create a key experiment on SemaFor, in collaboration with NVIDIA. The experiment tested the ability to detect images from Generative Adversarial Networks (GANs) without training data from the architecture, mimicking the threat posed by adversaries that could develop novel GAN architectures. SemaFor performers demonstrated the ability to detect images from StyleGAN3 with high accuracy---0.99+ area under receiver operating characteristic curve (AUC ROC)---and no training data from the architecture and no knowledge of the architecture~\cite{nvidia_research_projects_nvlabsstylegan3-detector_2023}. Crucially, this experiment was conducted prior to NVIDIA’s release of StyleGAN3, so there was no way that training information could have leaked. NVIDIA held the release of StyleGAN3 until the detectors were available and then both StyleGAN3 and the detectors were released publicly on the same day.

One particularly compelling capability from MediFor and SemaFor was the development of person-of-interest (POI) or soft-biometric models. These models learn person-specific facial motion patterns based on head pose and facial action units for expressions~\cite{agarwal_detecting_2020,agarwal_protecting_2019}. Trained on known good data (about an hour of video) for an individual, these models have demonstrated the ability to discriminate the real individual from impersonators, deepfakes, and impersonators + deepfakes. Such models provide a compelling defense for high-profile individuals that might be targets of attacks. Continuing research is necessary to extend the models to a broader range of viewpoints and to incorporate additional information such as gestures, body pose, and speech. Recently, DARPA has announced {\it The AI Forensics Open Research Challenge Evaluations (AI FORCE)} with
the aim of developing and evaluating techniques to mitigate the threats posed by state-of-the-art AI systems~\footnote{See \url{https://semanticforensics.com/ai-force/challenge-1}}.  This challenge will be very useful in evaluating
state-of-the-art deepfake detection techniques.


\subsection{Watermarking of GenAI Output}
\label{sec:watermarking}

\newcommand{\MES}[1]{\textcolor{blue}{Mehrdad: #1}}



As described below, a watermarking method has two components.
\begin{itemize}
\item An encoder ${\rm Embed}_{k_E} (M,p,w)$ where $k_E$ is a secret key, $M$ is the model, $p$ is user supplied input to the model (e.g. a prompt or instructions for editing a message), and $w$ is additional information (e.g. string to be embedded in a watermark). Some schemes might not use some parameters. For example, if a scheme does not use a secret key, then $k_E$ will not be used, and in some schemes $w$ might not be used. 
\item A decoder ${\rm Detect}_{k_D} (x,w)$ where $k_D$ is the key use for detection, $x$ is content, and $w$ is additional information. As usual, some schemes might not use certain parameters, such as $k_D$ and $w$. This method returns $1$ if $x$ is watermarked, and $0$ otherwise. 
\end{itemize}
Note that in secret key schemes, such as~\cite{neurocryptography,kuditipudi2023robust,christ2023undetectable}, $k_E = k_D = k$ and is kept secret. In publicly-verifiable schemes~\cite{fairoz}, $k_E$ is the secret key and $k_D$ is the public key. As we already stated, the precise parameters of ${\rm Embed}$ and ${\rm Detect}$ depend on the watermarking scheme. 

Watermarking methods can be categorized into {\it non-learning-based}~\cite{pereira2000robust,boney1996digital,kirchenbauer2023watermark} and {\it learning-based}~\cite{zhu2018hidden,liu2023detecting,abdelnabi2021adversarial}. The former manually design encoder and decoder; while the latter uses neural networks as encoder/decoder and trains them using deep-learning techniques. In the image and audio domains, non-learning-based watermarking methods~\cite{pereira2000robust,boney1996digital} have been studied for several decades, while learning-based watermarking methods~\cite{zhu2018hidden,liu2023detecting} were proposed in the last several years. In the text domain, both non-learning-based~\cite{kirchenbauer2023watermark} and learning-based~\cite{abdelnabi2021adversarial} methods were proposed in recent years. 

One unique advantage of learning-based watermarking is that they can leverage adversarial training~\cite{goodfellow2014explaining}, a standard technique to build robust machine learning systems, to enhance robustness against post-processing that aims to remove the watermark in watermarked content. The key idea of adversarial training in the context of watermarking is to introduce a post-processing layer between the encoder and decoder~\cite{zhu2018hidden}. When training the encoder and decoder, the post-processing layer manipulates the watermarked content produced by the encoder via post-processing operations such that the learnt decoder can still accurately decode the watermark from a post-processed watermarked content. 

We note that some methods embed the watermark and encoder into the parameters of a GenAI model such that its generated content intrinsically has the watermark embedded~\cite{fernandez2023stable}. The attacks on watermarking discussed below are also applicable in such settings, but we will not discuss them in great detail. 

Watermarks are not perfect and not robust to all adversarial manipulations. There are also some impossibility results that a ``perfectly robust" watermark might not be possible~\cite{zhang_watermarks_2023}.
In order to understand what are good use cases for watermarking, it is very important to understand the threat landscape for various watermarking schemes. We discuss the current state of the art of attacks on watermarking schemes. We acknowledge that our treatment is not complete, but meant to give a flavor of the type of attacks on watermarking schemes. 

\subsubsection{Current State-of-the-Art of Attacks on Watermarking}

\begin{description}

   \item[Common Post-processing.] Watermarked content often undergoes various common post-processing operations in non-adversarial settings. These operations, while possibly not malicious in intent, may inadvertently remove the watermark. For instance, common image post-processing operations include compression, resizing, cropping, and color adjustments; typical text post-processing involves paraphrasing, word insertion, word deletion, and structural modifications; and popular audio post-processing includes compression, filtering, and re-recording. 

   Non-learning-based watermarking methods are often not even robust against common post-processing, e.g., JPEG compression removes the image watermark inserted by non-learning-based image watermarking methods~\cite{jiang2023evading} and paraphrasing removes the text watermark inserted by non-learning-based text watermarking methods~\cite{krishna2023paraphrasing, sadasivan2023can}. However, prior studies~\cite{zhu2018hidden,jiang2023evading} showed that learning-based image watermarking methods can be robust against common post-processing because they can leverage adversarial training. We expect learning-based text watermarking to be also more robust against common post-processing than non-learning-based ones due to adversarial training, though no prior studies have explored this. 

   \item[Diffusion Purification Attack.] Diffusion purification involves the process of passing data through forward and backward diffusion model steps for a specified number of diffusion steps ($t$). To elaborate, diffusion purification introduces Gaussian noise to the content and then utilizes denoising diffusion models to undo the Gaussian noise in order to get an output that is similar to the input. The parameter $t$ determines the degree of similarity between the output and the input. This technique has been used as a defense against adversarial attacks (Nie et al.~\cite{nie_diffusion_2022}), and also to remove watermarks from images (Saberi et al.~\cite{saberi_robustness_2024}, Zhao et al.~\cite{zhao_generative_2023}). Saberi et al.~\cite{saberi_robustness_2024} proposed a theoretical guarantee that diffusion purification can successfully attack watermarking techniques that introduce small perturbations to the content in order to watermark it (i.e., imperceptible watermarks).

   \item[Adversarial Post-processing.] Adversarial post-processing represents a strategic manipulation by attackers aimed at removing watermarks from content without compromising its quality. This section delves into the application of adversarial examples, originally introduced by Goodfellow et al.~\cite{goodfellow2014explaining}, to the domain of watermarking, focusing on  white-box, black-box, and no-box settings. 

   In the case where the attacker has white-box, black-box, or no-box access to the watermarking decoder, Jiang et al.~\cite{jiang2023evading} extended adversarial examples to image watermarks. Their research demonstrated that by introducing a small, human-imperceptible perturbation to a watermarked image, an attacker can effectively remove the watermark with theoretical guarantees. When the attacker lacks white-box access, they can find the perturbation by repeatedly querying the detection API. It's important to note that these white-box and black-box attacks do not require training any surrogate model. Additionally, Jiang et al.~\cite{jiang2023evading} developed a no-box attack by training a surrogate watermarking decoder. Hu et al.~\cite{hu2024transfer} further extended this approach with a transfer attack that involves training multiple surrogate watermarking decoders. This attack generates perturbations by aggregating outputs from multiple surrogate watermarking decoders.
   
   Besides, Zhang et al.~\cite{zhang_watermarks_2023} demonstrated the theoretical impossibility of creating robust watermarks against adversarial attacks that have black-box access to the model. Their analysis relies on two conceptual oracles: a quality oracle assessing output quality and similarity to the original data, and a perturbation oracle that can alter data while maintaining acceptable quality. 
   
   Moreover, Saberi et al.~\cite{saberi_robustness_2024} have demonstrated that surrogate model adversarial attacks can effectively compromise image watermarking techniques, especially those employing high-perturbation watermarks (i.e., a family of watermarking techniques that significantly alter the original data, and usually have higher robustness to non-adversarial attacks \cite{saberi_robustness_2024}). These attacks involve training a surrogate model to mimic the watermark decoder using a collection of watermarked images, eliminating the need for direct access to the actual watermark decoder. The trained surrogate decoder can subsequently be employed to apply adversarial perturbations to the data. Furthermore, An et al.~\cite{an2024benchmarking} have developed an extensive benchmark to evaluate the robustness of watermarks, offering valuable insights into their effectiveness.

   The literature on attacking watermarking schemes mostly tackle text and image modalities. Modalities, such as audio and video, have received scant attention. For example, existing attacks on text and audio watermarking are mostly based on common post-processing. It is an interesting future work to explore adversarial-example-based attacks to text, audio, and video watermarks. 
   
\end{description}

\subsubsection{Plausible Use Cases for Watermarking}


Drawing on insights from recent studies~\citep{jiang2023evading, hu2024transfer, saberi_robustness_2024, an2024benchmarking, sadasivan2023can} and impossibility results~\cite{zhang_watermarks_2023} 
regarding the robustness of watermarking techniques, it becomes apparent that no existing watermarking method is universally effective under all circumstances and against a powerful adversary. This is particularly true within strict regimes such as TPR@1\%FPR (true positive rate at 1\% false positive rate), where the reliability of watermarks significantly diminishes, even against non-adversarial attacks accessible to attackers at minimal costs. Nonetheless, the literature does not entirely rule out the possibility of developing reliable watermarking schemes in the future.

There may be scenarios where existing watermarking techniques would still be beneficial. This can include novel applications of watermarks for a range of downstream tasks. In this section, we explore some possible uses for watermarks, taking into account the limitations identified in prior research.
We acknowledge this is not an exhaustive list of plausible use cases for watermarking. More investigation is needed to explore plausible use cases
for watermarking.

\begin{description}
    \item[Non-Critical Use Cases.] For tasks where having high robustness against attackers and manipulators is not critical (e.g., watermarking personal data and photos, just to discourage their sharing and copying, and avoiding data feedback loops during model training).
     
    \item[Adversary-Restricted Environments.] Settings where access for adversarial attacks (either attacks with decoder access, or surrogate model attacks), are inherently limited. This can also include settings where some common post-processing attacks might not be aligned with the attacker's objectives. For instance, consider an image watermarking technique that is not robust to flipping the input images. If the watermark is applied to images of items such as documents or posters, flipping these images would yield impractical results, thus negating the attack's effectiveness.
    
    \item[Verifying Content and its Ownership.] In this case, attacks that can remove the watermark from data are not a concern. Instead, our only concern is about the spoofing error of the watermark being low.
    Spoofing corresponds to an unwatermarked content (say a legal document or harmful content) and creating a semantically equivalent content that is watermarked (note that this is similar to forgery attacks on digital signature schemes). Note that publicly verifiable schemes, such as~\cite{fairoz}, provably thwart spoofing attacks.
    These watermarks can be used to verify legal documents and other sensitive data, or be used to verify the ownership of the data (the original owner will provide the watermark key to prove ownership). This can also be done without altering the content using extra metadata.
    
\end{description}

\section{Gaps between Policy Goals and Technology Capabilities}
\label{sec:gaps}


\paragraph{Alignment.}

The most significant gap is likely in the GenAI alignment space, where technical capabilities to align models so that they satisfy policy goals are limited and lagging behind the development of other GenAI capabilities. Ensuring that models are safe and reliable to use, do not produce harmful or deceitful outputs, exhibit robustness to adversarial inputs, and do not pose societal threats is an ongoing endeavour. The field lacks a clear metric for alignment (beyond the results of loss minimization when preference tuning and preference optimization), and thus at this point it is not feasible to rank models in terms of alignment.

On the policy side, different governments emphasize different aspects of alignment in their regulation (as we observe in \autoref{sec:policy}) thus making it likely that models aligned for one country may be banned as unsafe in another. This poses challenges in terms of cost and effort for model producers, as they need to custom-align their deployed models for various countries. Furthermore, alignment can affect model performance and vice versa~\cite{qi_fine-tuning_2023}, and overly cautious alignment can reduce model utility, as the aligned models can refuse to produce outputs for prompts that ``superficially resemble unsafe ones.''~\cite{bianchi_safety-tuned_2023}

\paragraph{Liability.}

Regulations surrounding GenAI are built upon the foundations of clear attribution, predictable behavior and transparency but current technology is struggling to meet these requirements. A major issue arises from the ``black box'' nature of GenAI models, particularly those rooted in deep learning where tracing the origins of harmful outputs to specific actors, whether it be code or training data, is a significant challenge. This opacity obfuscates liability as it is nearly impossible to pinpoint the exact cause of problematic generations. Additionally the stochastic nature of GenAI introduces unpredictability, as models might not always adhere to set guidelines or avoid harmful outputs. This lack of determinism makes assigning blame difficult when the same input doesn't consistently produce the same output.

The distribution of responsibilities in the GenAI landscape further compounds the issue. From data providers to model trainers and end-users, multiple parties are involved which makes determining liability complex---if a model is trained on biased data, who bears the responsibility, the data provider or the model trainer? To bridge this gap, solutions such as {\it explainable-AI} may be considered to make these systems more transparent, coupled with industry-wide technical standards and adaptive regulatory frameworks which can evolve with technology. 


\paragraph{Watermarking Limitations and Unachievable Regulatory Mandates.}

Multiple legal mandates are in place to regulate the labeling of GenAI outputs. For instance, the EU AI Act stipulates that GenAI must adhere to transparency protocols, which include labeling AI-generated content and preventing the creation of illegal content~\cite{EU-AI-Act}. Similarly, the White House's Executive Order on the Safe, Secure, and Trustworthy Development and Use of Artificial Intelligence emphasizes the use of watermarking as a key technique for labeling AI-generated content~\cite{the_white_house_executive_2023}. 

On the one hand, legal regulations should align with the capabilities and limitations of current watermarking technology. As mentioned before, watermarking techniques exist for various modalities, such as text, images, video, and audio. There are two major dimensions along which the watermarking techniques differ – quality (how much watermarking degrades the quality of the content) and robustness (how easy it is for an adversary to bypass the detection scheme associated with the watermarking scheme). The state of the art of the watermarking schemes for various modalities differ along these two dimensions. Therefore, we contend that legislation needs to be aware of the state of the art of watermarking for various modalities.
On the other hand, policies and laws can serve as effective instruments to support the technical landscape of
watermarking. Currently, regulations primarily target GenAI service providers. These regulations could be broadened
to also govern the behaviors of users and attackers. 
Policymakers could consider the legal implications if someone attempts to remove watermarks from AI-generated content.
This extension would provide an additional layer of protection against misuse and manipulation of GenAI.

\section{Future Directions for Regulators and Technologists}
\label{sec:future}


At the workshop, the participants identified several areas where regulators and technologists can work together to build a path to safe GenAI.

\paragraph{Adjust Speed of Regulation to Risk of Use Cases.}




To match the rapid pace of GenAI innovation, regulatory frameworks need major overhauls to develop increased agility. Traditional policy development cycles, often spanning years, are fundamentally incompatible with the breakneck speed of technological advancement. Instead of these drawn-out processes, we need solutions like expert working groups empowered to make rapid recommendations based on the latest research. Additionally, regulatory sandboxes would allow experimentation with new rulesets in controlled environments, minimizing the risk of unintended consequences.  Finally, ongoing collaboration between technologists and policymakers is crucial – this will streamline the process of translating highly technical developments into sound, adaptable regulations that minimize the lag between technological breakthroughs and their safe integration into society.

Overly broad regulation risks stifling the very innovation policymakers seek to govern. Current regulation aims for highly desirable guarantees from GenAI, including privacy, fairness, interpretability, but only defines them with broad strokes.  GenAI is an incredibly diverse field, with applications ranging from creative image generation to life-saving drug discovery, and thus one-size-fits-all approaches will not work. Instead of blanket pronouncements that might inadvertently hinder beneficial research directions, regulation needs a nuanced understanding of the specific risks and benefits associated with each use case. Frameworks outlining overarching principles such as fairness, transparency, and accountability are essential and most useful when adjusted to specific domains of application and their corresponding risk levels.  This approach offers flexibility to developers while ensuring ethical development, fostering responsible innovation without unnecessarily hampering progress.




\paragraph{Enable Interdisciplinary Discussions to Break Out of Silos.}




While the tendency in (machine learning) research is to seek out specialized communities for efficiency and focus, the rapid evolution of GenAI underscores the critical importance of enabling robust interdisciplinary research to fully realize the technology's potential and mitigate its risks. We need to foster forums that encourage discussions and collaborations across disparate specializations, bringing together a diverse range of perspectives to chart the course of this transformative technology. The current momentum demands that we think proactively about its long-term trajectory, ensuring that the path we forge aligns with our broader goals and values. Left unchecked, the siloed nature of research, coupled with evolving challenges in areas like regulation, could lead to fragmented solutions that address only narrow aspects of the problem, without the broader perspective that interdisciplinary research offers.  To incentivize this crucial broadening, we need to re-examine our systems for publishing and grant allocation, ensuring they promote fresh perspectives and collaboration across fields, rather than inadvertently reinforcing the tendency toward repetitive work within existing silos.

\paragraph{Support Sharing of Lessons from Failures.}



GenAI models excel in generating human-quality responses but often fail to identify and address their own shortcomings.  The publication of research results, including attacks and defenses on GenAI, progresses apace but often lacks the real-world specificity and relevance of lessons from GenAI deployments.  This underscores the critical need for widespread information sharing of security and privacy lessons derived from GenAI failures, spanning both academic and commercial settings. Collaboration across these sectors is essential because the vulnerabilities of one model can directly inform preventative measures and proactive protections for others operating in similar domains. We can draw a direct parallel with cybersecurity, where information sharing is a cornerstone of threat intelligence. It took cybersecurity many years to reach a significant level of coordination and sharing, starting in 1988 with the creation of the Computer Emergency Response Team Coordination Center (CERT/CC) to maintain a repository of vulnerability information, and as recently as 2021 moving to ``continuously exchange, enrich, and act on cybersecurity information'' through the Joint Cyber Defense Collaborative (JCDC) of the US government. Just as cybersecurity professionals share insights on malware, phishing scams, and network vulnerabilities, the GenAI community must establish robust channels for exchanging knowledge on model failures, adversarial attacks, and potential societal harms. By understanding why and how models falter, we can collectively develop better safeguards, mitigate biases, and build a more responsible and resilient GenAI ecosystem for the future.

\paragraph{Encourage the Research and Development of Out-of-model Safety Guardrails.}

There is a tremendous focus on making GenAI models safe (based on varying definitions of safety), while maintaining the models' accuracy, usefulness, speed, and emerging capabilities. This directs all of the efforts of the technology community towards a narrow approach of creating models that are as close to perfect as possible, which appears to be both theoretically and practically impossible (as we discussed in \autoref{sec:alignment}). An avenue that is less explored but holds a lot of promise is to consider the safety, security, and privacy of the GenAI-based system, not just the model itself. This means going beyond chatbot-style systems and considering that the vast majority of real-world applications are likely to deploy GenAI models in a larger system with specific requirements, interfaces, and goals (in contrast to the unbounded operation of chatbots). Topics such as the secure sequential and parallel composition of GenAI-based systems, layered security for multi-agent systems, security uses of watermarked GenAI outputs, and model explainability for security and privacy can advance the state of safety for GenAI-based systems. Initial research along these directions identified the need for coupling the ML model with a safety specification, a verifier, and a world model, whose practical realizations require new technical advances~\cite{dalrymple_towards_2024}.


\section{Conclusions}
\label{sec:conclusions}


We summarized the discussions at a workshop held in October 2023 co-organized by Google, University of Wisconsin--Madison, and Stanford University, on the topics of GenAI policy and safety. We hope this underscores the urgent need for a dynamic and nuanced approach to GenAI regulation, adapting to the pace of rapid technological advancements and their varied implications across different sectors of society. On the policy side we highlighted numerous efforts to regulate GenAI safety, resulting in a patchwork of mandates with various emphases. On the technology side three technologies were considered, safety alignment, model watermarking, and model interpretability, though all of them currently have limitations, some of them fundamental.

The collaborative efforts at the international level, such as those detailed from the G7 and other bodies, illustrate the beginning of a more cohesive and comprehensive approach to GenAI governance, but much work remains to ensure these efforts are effective and inclusive. Gaps arising from policy requirements mismatch against technical capabilities must be addressed even when technical means are not available, and we drew attention to risk-based system designs, where policy--technology gaps are tackled head on instead of waiting for future technology improvements. Finally, we put forth several directions for both regulators and technologists to consider when seeking guardrails and safety in GenAI-based systems.


\section*{Acknowledgements}
\label{sec:ack}


We thank all of the speakers and participants to the workshop for their insightful contributions and Google, NIST, DARPA, CMU, University of Maryland College Park, University of Virginia, Georgetown’s Center for Security and Emerging Technology, Harvard, Duke, and many others for the logistical support.


\bibliographystyle{plainurl}
\bibliography{references, yanjunqi-model-eval, refs-Neil-Zhengyuan, refs-Mehrdad-Watermark, somesh}

\end{document}